\documentclass[namedreferences]{solarphysics}

\usepackage[hyperref,optionalrh]{spr-sola-addons} 
\usepackage{graphicx}        
\usepackage{color}           
\usepackage{breakurl}        

\usepackage{sistyle}
\usepackage[clockwise, figuresleft]{rotating}

\newcommand{\aap}{    {\it Astron. Astrophys.}}

\newcommand{\apj}{    {\it Astrophys. J.}}
\newcommand{\apjl}{   {\it Astrophys. J. Lett.}}
\newcommand{\apjs}{   {\it Astrophys. J. Supplem.}}

\newcommand{\grl}{    {\it Geophys. Res. Lett.}}

\newcommand{\jgr}{    {\it J. Geophys. Res.}}

\newcommand{\solphys}{{\it Solar Phys.}}
 
\newcommand{\ssr}{    {\it Space Sci. Rev.}} 
\chardef\us=`\_

\begin{document}

\begin{article}

\begin{opening}

\title{CME -- HSS interaction and characteristics tracked from Sun to Earth}

%
\author[addressref={graz},corref,email={stephan.heinemann@hmail.at}]{\inits{S.G.}\fnm{Stephan G. }\lnm{Heinemann}\orcid{0000-0002-2655-2108}}
\author[addressref={graz},corref,email={}]{\inits{M.}\fnm{Manuela }\lnm{Temmer}\orcid{0000-0003-4867-7558}}
\author[addressref={unh},corref,email={}]{\inits{C.}\fnm{Charles J. }\lnm{Farrugia}\orcid{}}
\author[addressref={graz},corref,email={}]{\inits{K.}\fnm{Karin }\lnm{Dissauer}\orcid{0000-0001-5661-9759}}
\author[addressref={nasa,cua},corref,email={}]{\inits{C.}\fnm{Christina }\lnm{Kay}\orcid{0000-0002-2827-6012}}
\author[addressref={mp},corref,email={}]{\inits{T.}\fnm{Thomas }\lnm{Wiegelmann}\orcid{}}
\author[addressref={graz},corref,email={}]{\inits{M.}\fnm{Mateja }\lnm{Dumbovi\'c}\orcid{0000-0002-8680-8267}}
\author[addressref={graz,kso},corref,email={}]{\inits{A.}\fnm{Astrid M. }\lnm{Veronig}\orcid{0000-0003-2073-002X}}
\author[addressref={skoltech},corref,email={}]{\inits{A.}\fnm{Tatiana }\lnm{Podladchikova}\orcid{}}
\author[addressref={graz},corref,email={}]{\inits{S. J.}\fnm{Stefan J. }\lnm{Hofmeister}\orcid{0000-0001-7662-1960}}
\author[addressref={unh},corref,email={}]{\inits{N.}\fnm{No\'e }\lnm{Lugaz}\orcid{0000-0002-1890-6156}}
\author[addressref={spain},corref,email={}]{\inits{N.}\fnm{Fernando }\lnm{Carcaboso}\orcid{0000-0003-1758-6194}}


%

\address[id={graz}]{University of Graz, Institute of Physics, Universit\"atsplatz 5, 8010 Graz, Austria }
\address[id={unh}]{University of New Hampshire,Institute for the Study of
Earth, Oceans, and Space, Morse Hall, 8 College Road, Durham, NH 03824-3525, USA}
 \address[id={nasa}]{Solar Physics Laboratory, NASA Goddard Space Flight Center, Greenbelt, MD, USA}
 \address[id={cua}]{Dept. of Physics, The Catholic University of America, Washington DC, USA}
 \address[id={mp}]{Max-Planck-Institut f\"ur Sonnensystemforschung, Justus-von-Liebig-Weg 3, 37077 G\"ottingen, Germany}
 \address[id={kso}]{Kanzelh\"ohe Observatory for Solar and Environmental Research, University of Graz, Austria}
 \address[id={skoltech}]{Skolkovo Institute of Science and Technology Skolkovo Innovation Center, Building 3 Moscow 143026, Russia}
 \address[id={spain}]{Dpto. de F\'isica y Matem\'aticas, Universidad de Alcal\'a, 28805 Alcal\'a de Henares, Madrid, Spain}
\begin{abstract}
In a thorough study, we investigate the origin of a remarkable plasma and magnetic field configuration observed \textit{in situ} on June 22, 2011 near L1, which appears to be a magnetic ejecta (ME) and a shock signature engulfed by a solar wind high--speed stream (HSS). We identify the signatures as an Earth-directed coronal mass ejection (CME), associated with a C7.7 flare on June 21, 2011, and its interaction with a HSS, which emanates from a coronal hole (CH) close to the launch site of the CME. The results indicate that the major interaction between the CME and the HSS starts at a height of $1.3R_{\odot}$ up to 3$R_{\odot}$. Over that distance range, the CME undergoes a strong north--eastward deflection of at least $30^{\circ}$ due to the open magnetic field configuration of the CH. We perform a comprehensive analysis for the CME--HSS event using multi--viewpoint data (from the \textit{Solar TErrestrial RElations Observatories}, the \textit{Solar and Heliospheric Observatory} and the \textit{Solar Dynamics Observatory}), and combined modeling efforts (nonlinear force--free field modeling, \textit{Graduated Cylindrical Shell} CME modeling, and the \textit{Forecasting a CME’s Altered Trajectory -- ForeCAT} model). We aim at better understanding its early evolution and interaction process as well as its interplanetary propagation and related \textit{in situ} signatures, and finally the resulting impact on the Earth's magnetosphere.
\end{abstract}

%

\end{opening}

%
 \section{Introduction}
Coronal mass ejections (CME) frequently pass over Earth at an average rate of $1-2$ events per month but with significant variations throughout the solar cycle \citep{2010richardson_RC-list}. They are the major cause of strong geomagnetic effects, especially during solar maximum \citep[\textit{e.g.},][]{1997farrugia,liu14}. Besides CMEs, stream interaction regions (SIR) and, if persistent for several solar rotations so--called co--rotating interaction regions (CIR), structure interplanetary space. It is the interaction between solar wind high speed streams (HSS), emanating from coronal holes (CH), and the slow solar wind ahead that forms compression regions, shocks and rarefaction regions, causing recurrent geomagnetic effects on Earth (see, \textit{e.g.}, \citealt{alves06,verbanac11,2017vrshnak,2018yermolaev,2018richardson}). While HSSs pose a continuous outflow, CMEs abruptly disrupt the rather steady solar wind structure, causing deviations (preconditioning) from the quiet solar wind conditions over the duration of several days \citep{2017temmer-precond}. 

The interaction of CMEs with the solar wind, especially HSSs, may significantly change the CME properties en route through the heliosphere. The embedded flux rope may deform, kink or rotate  \citep{2004manchester,2004riley,2006wangY,2008Yurchyshyn,2013Isavnin}, erode due to reconnection \citep{2006Dasso,2012ruffenach,2014lavraud,2015ruffenach,2018wangY}, be deflected \citep{2004wangY,2005manchester,2013kay,2014wangY,2015kay,2016wangY,2019zhuang} and may be related to increased turbulence in the sheath region  \citep{2015lugaz,kilpua17b}. Fast solar wind may also cause the CME to speed up, hence, shortening the propagation time between Sun and Earth. \cite{2012Nieves-Chinchilla} investigated in detail the rotation of a CME as it propagates through the heliosphere. It is understood that changes in CME properties differ strongly for processes taking place already low in the corona compared to those happening in interplanetary space \citep{2014wang,2016winslow}. All these effects alter the initial CME properties observed close to the Sun making predictions of arrival time and geoeffectiveness a complex endeavor \citep{2018richardson}. This also shows that the ambient solar wind plays an important role for CME propagation, and that it is necessary to study and understand the interaction processes already from its source on the Sun and related 1 au signatures.

Previous studies of CME--HSS interaction events often focused on HSSs catching up with a CME deforming and compressing it \citep{2016winslow, 2018wen}. In this study the CME appears to be propagating behind the SIR within the HSS, giving rise to a number of peculiarities in the observed \textit{in situ} signatures. By studying in detail this CME--HSS interaction event we aim to unravel the complex physical processes related to a CME propagating in a HSS starting from the CME eruption site close to the HSS related CH.  To do so, we investigate the Sun--Earth chain of a distinct and well--observed CME--HSS interaction event combining remote sensing observations, \textit{in situ} measurements at 1 au and modelling efforts. We further investigate the effects on the Earth's magnetosphere as a consequence of the CME--HSS interaction.

\section{Motivation}
Starting on June 22, 2011, we observed an intriguing configuration of plasma and magnetic field \textit{in situ} signatures near L1. We find a SIR signature with a clear stream interface (SI) followed by a shock within the HSS on June 23, 2011 (Figure~\ref{fig:in-situ}). After a short standoff distance, the magnetic ejecta (ME) can be identified propagating with the same speed as the ambient solar wind of the HSS. This gives rise to two major questions which we address in this study:
\begin{itemize}
\item \textit{How can these unique shock characteristics be explained?} Usually, a shock within a high speed environment, like a HSS, would either quickly propagate through or would dissipate in case it is not driven. The intuitive solution would be that the shock is driven by the ejecta.
\item \textit{Is the shock driven?} Judging from the short standoff distance between the shock and the ejecta and that the shock signature is found in the middle of the HSS, one would suspect so. However, the speed of the ejecta is equal to the speed of the HSS, but the speed of the shock is higher, which indicates that it is {\em not} driven at that distance.
\end{itemize}

To answer these questions we track the CME from 1 au back through the interplanetary space to its origin on the solar disk and examine what processes may produce such \textit{in situ} signatures and what geomagnetic effects they cause. The paper is structured as follows: In Section~\ref{sec:in-situ} the \textit{in situ} signatures are investigated. In Section~\ref{sec:remote+modelling} the CME launch and propagation is analyzed using remote sensing data and various modelling efforts. Section~\ref{sec:geomag} presents the resulting geomagnetic effects. In Section~\ref{sec:disc} the results are discussed and summarized in Section~\ref{sec:summary}.

 \section{\textit{In situ} Signatures}\label{sec:in-situ}
Using OMNI\footnote{https://omniweb.gsfc.nasa.gov/} data from the \textit{Advanced Composition Explorer} (ACE; \citealt{1998stone_ACE}) and the \textit{Global Geospace Science Wind} satellite \citep{1995acuna_GSS}, we investigate the properties of the solar wind and CME plasma at a distance of about 1 au.

\begin{figure} 
 \centerline{\includegraphics[width=0.9\textwidth,clip=]{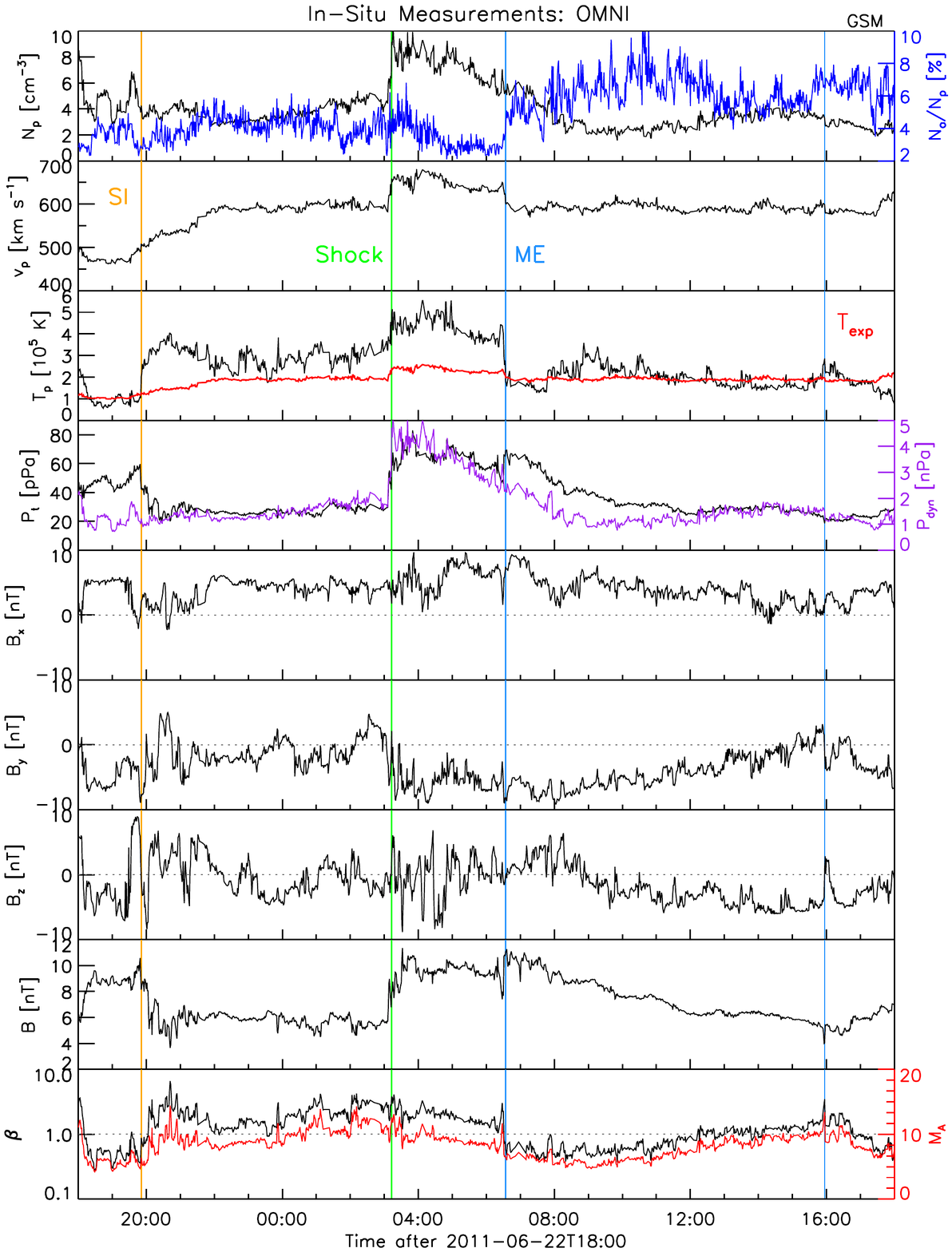}}
 \caption{\textit{in situ} measured \textit{WIND} and ACE data from the OMNI database of a 24h time interval during the HSS and CME arrival starting at June 22, 2011 18UT. From \textit{top to bottom}: Proton density (\textit{black}) with the $\alpha$--particle ratio overlayed (\textit{blue}); proton velocity; proton temperature (\textit{black}) with the expected temperature in red; total perpendicular pressure (\textit{black}) and the dynamic pressure (\textit{purple}); magnetic field components (panel 5--8); plasma--$\beta$ (\textit{black}) and the \'Alfvenic Mach number (\textit{red}). The vertical guidelines represent the start times of the stream interface (\textit{orange}), the shock (\textit{green}) as well as the interval identified as the magnetic ejecta (\textit{blue lines}) as derived from the data in this figure.}\label{fig:in-situ}
 \end{figure}

Figure~\ref{fig:in-situ} shows a 24--hr period from 18 UT, June 22 to 18 UT, June 23, 2011. From top to bottom, we show the proton number density and the alpha--to--proton number density ratio, the bulk speed, the temperature and expected proton temperature for normal solar wind expansion \citep{1987lopez}, the total perpendicular pressure \citep{2005russell} and dynamic pressure including the $\alpha$ particles, the components of the magnetic field in GSM coordinates, the total magnetic field, and the proton $\beta$ and the Alfv\`en Mach number. 
From the proton bulk speed one can identify a slow stream followed by a fast one. The stream interface occurs at around 20 UT, June 22 shown by the orange vertical guideline. The SI was identified based on the sharp drop of the density compression and the sharp rise in the temperature as shown in Figure~\ref{fig:in-situ}. Additionally we observe a gradient in the flow, where the east--west flow changes sign (not shown). The following high speed solar wind stream is interrupted at 03:06 UT by simultaneous sharp rises in the density, temperature, speed and magnetic field, corresponding to a shock structure. Its strength is moderate, with a density ratio of $\sim2$ and \textit{in situ} a field compression of $\sim1.5$. By using the Coplanarity Theorem \citep{1966colburn, 1972Abraham-Shrauner,1976Abraham-Shrauner}, which assumes that the magnetic field on both sides of the discontinuity and the shock normal all lie in the same plane, we calculate the shock normal. The shock normal speed we derive using the mass conservation equation across the discontinuity \citep[\textit{e.g.}, see][and references therein]{1998Paschmann+Daly}. Using the magnetic coplanarity \citep{1966colburn} we calculate the shock normal velocity to be $\sim 710$ km s$^{-1}$. The angle between the upstream magnetic field and the shock normal ($\theta_{Bu,n}$) is $\sim 30^{\circ}$ which makes it a quasi--parallel shock. When using the mixed mode coplanarity \citep{1972Abraham-Shrauner} to calculate the shock normal, we derive $v_{shock} \sim678$ km s$^{-1}$ with $\theta_{Bu,n}  \sim 34^{\circ}$.  Behind the shock the elevated high speed continues for $\sim~3.75$ hrs. Then a structure with high magnetic field strength, low proton temperature, low beta and Alfv\'en Mach number, and increased alpha--to--proton number density ratio is present.

\begin{figure} 
 \centerline{\includegraphics[width=1\textwidth,clip=]{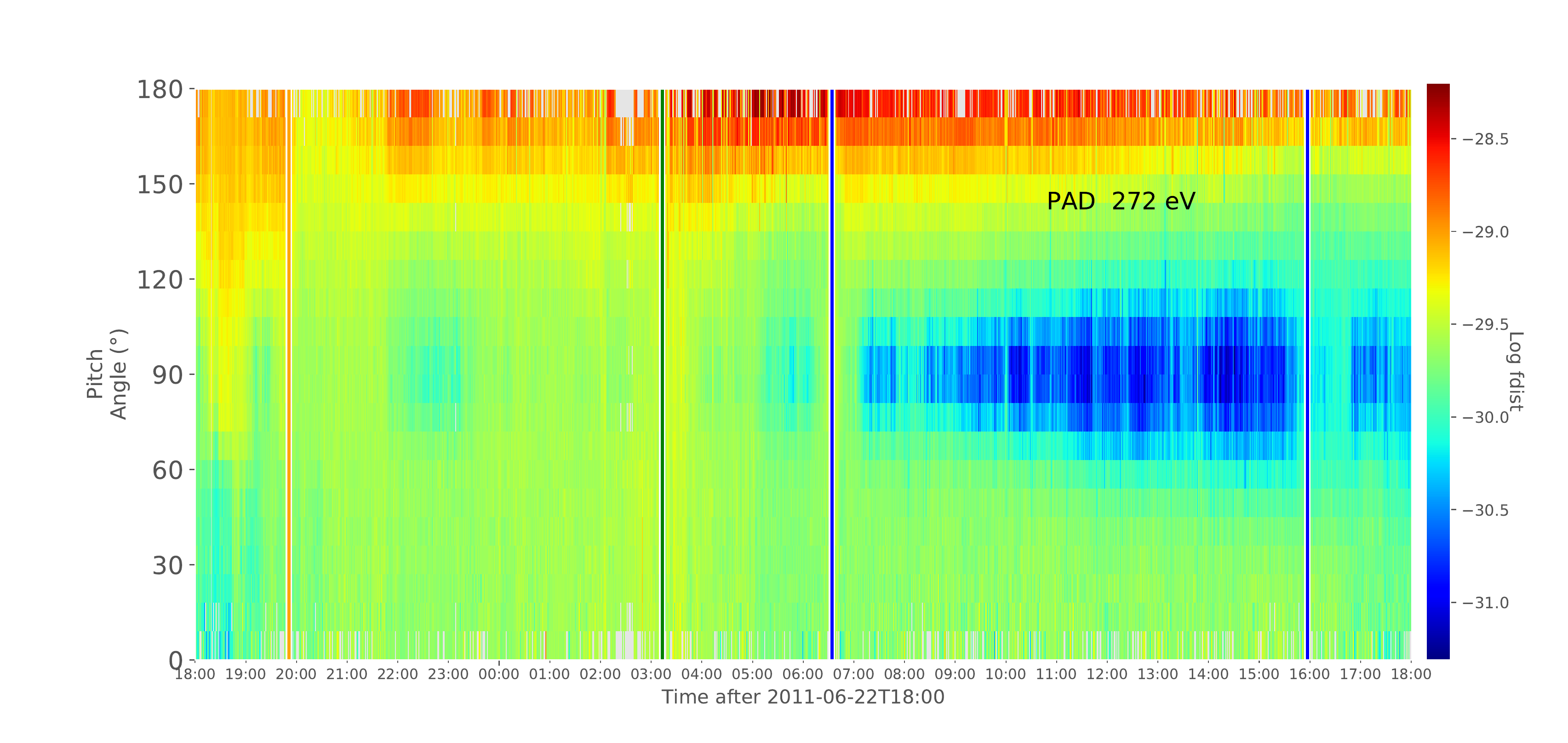}}
 \caption{Suprathermal electrons pitch--angle distribution observed by ACE for the 272 eV energy channel, time--shifted to match the OMNI data. A bidirectional distribution can be observed during the transit of the magnetic ejecta. The vertical guidelines are the same as shown in Figure~\ref{fig:in-situ}.}\label{fig:supra}
 \end{figure}

Figure~\ref{fig:supra} presents the pitch angle distribution of suprathermal electrons in the ACE $272$~eV channel showing isotropic flux during the shock arrival and the transition from the shock--sheath to the magnetic ejecta, which is expected. In this interval, we identify as the magnetic ejecta the period during which the electrons show bidirectionality, suggesting a closed magnetic structure (\textit{e.g.}, \citealt{1974montgomery,1987gosling_b,2018carcaboso}). This strongly supports the interpretation of a flux rope within the HSS.

We note that the bulk speed before the shock and after the shock--sheath is the same, which suggests that the magnetic ejecta is currently not driving the shock. This can be interpreted as follows: the CME is embedded and dragged along with the co--rotating interaction region (CIR), which is an intriguing feature.

 \section{Remote Observations \& Modelling}\label{sec:remote+modelling}
\begin{figure} 
  \centerline{\includegraphics[width=1\textwidth,clip=]{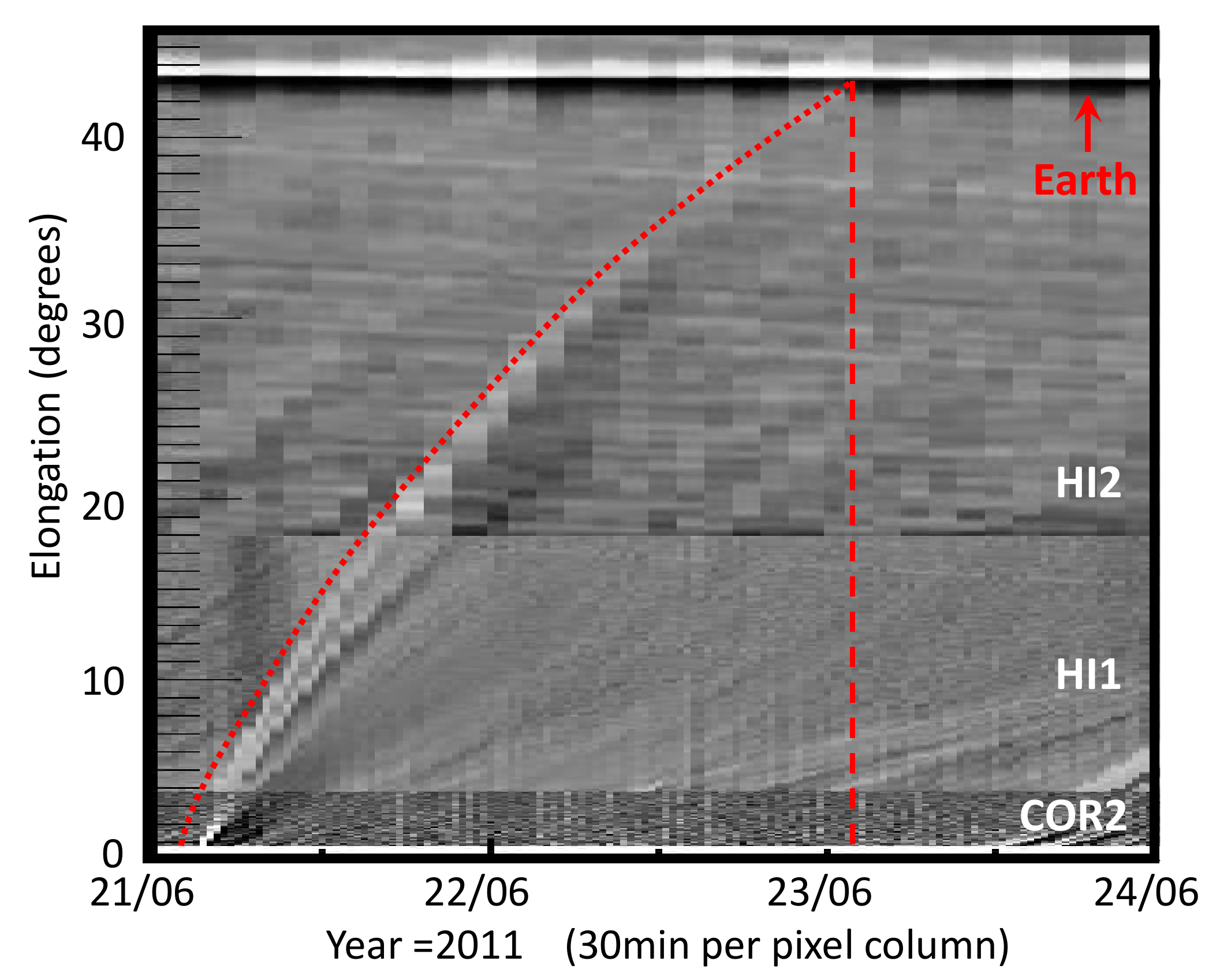}}
  \caption{J--map of the ICME as seen in COR2, HI1 and HI2. The ICME kinematic is marked by the \textit{red dotted line} and \textit{in situ} arrival of the shock by the \textit{red dashed line}. This feature is manually tracked back to the corresponding CME launch time at the Sun around 02 UT on June 21, 2011.}\label{fig:jmap}
  \end{figure}

\begin{figure} 
 \centerline{\includegraphics[width=1\textwidth,clip=]{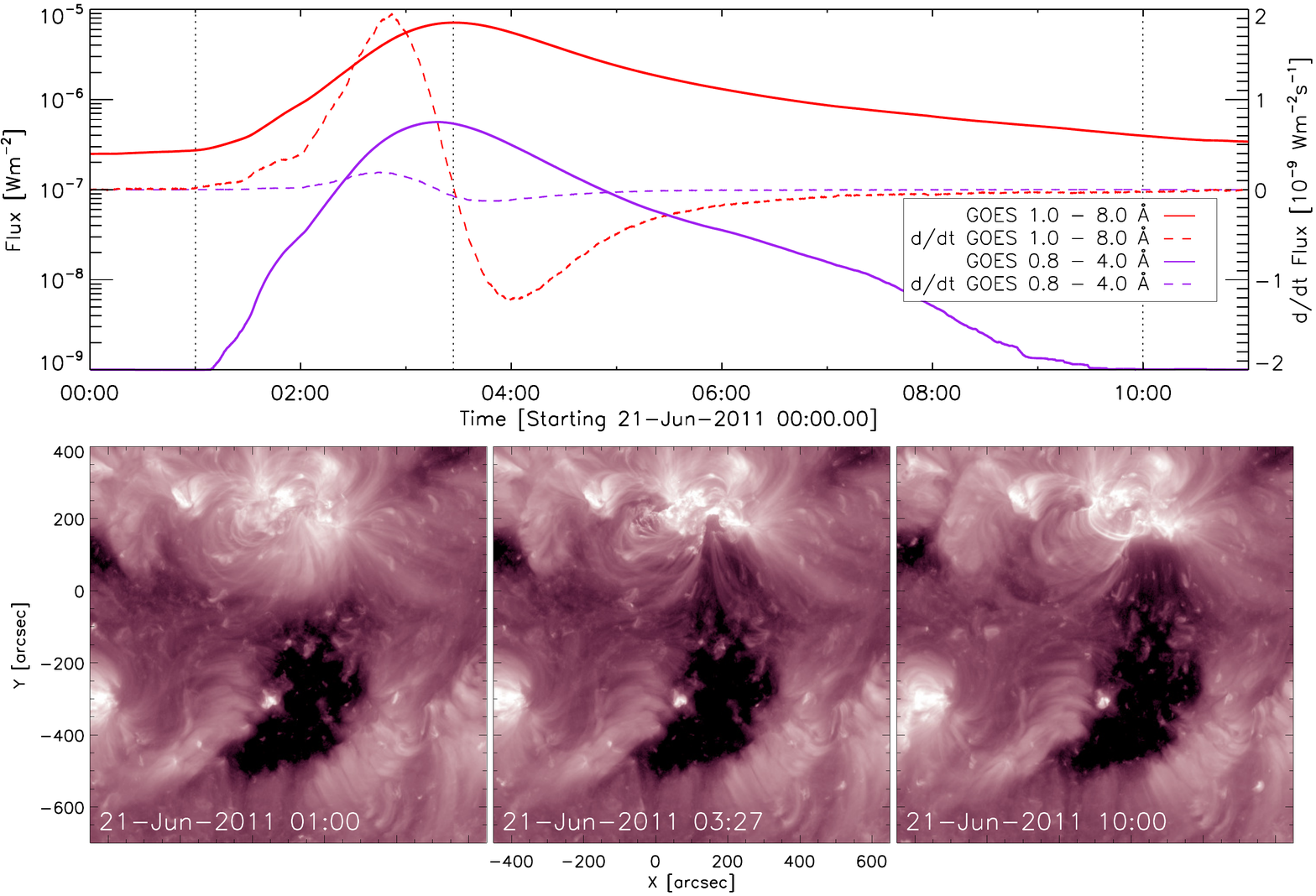}}
 \caption{\textit{Top}: GOES soft X--ray fluxes and time derivative of the observed C7.7 flare. \textit{Bottom}: AIA/SDO 211\AA\ images with the field of view centered around the source AR (N17/W12) and the CH located to the south. The image recording times are marked as the \textit{dotted vertical lines} in the  GOES soft X--ray profile and represent the configuration of the solar corona before, during and after the flare.}\label{fig:goes}
 \end{figure}

Using white--light as well as extreme ultra--violet (EUV) observations obtained by the \textit{Sun Earth Connection Coronal and Heliospheric Investigation} (SECCHI; \citealt{2008howard_SECCHI}) suite on board the \textit{Solar TErrestrial RElations Observatories} (STEREO--A/B; \citealt{2008kaiser_STEREO}) we are able to track the CME back to its origin on the solar disk. Figure~\ref{fig:jmap} shows the propagation of the CME along the equatorial cut as seen by STEREO--A's \textit{Heliospheric Imager} 1 and 2 (HI1, HI2) as well as its coronagraph (COR2) obtained in form of a J--map. We estimate the launch time at the Sun to be around June 21, 2011 02 UT. This date coincides very well with an observed gradual C7.7 flare from the active region (AR) 11236 (location: N17/W12) starting at 01:22 UT on June 21, 2011. The flare can be associated with an Earth directed halo CME whose signatures are subsequently observed by the \textit{Solar Dynamics Observatory} (SDO; \citealt{2012pesnell_SDO}), the \textit{Solar and Heliospheric Observatory} (SOHO; \citealt{1995soho}) and both STEREOs. A medium--size coronal hole (CH) is located in the south--west of the AR that we associate with the CIR observed \textit{in situ} near 1 au on June 22, 2011. Figure~\ref{fig:goes} shows the GOES soft X--ray flux around the time of the flare (top), which features a very gradual increase over nearly $2$ hrs before the peak in the soft X--ray flux ($1-8$~\AA) of $7.7 \cdot 10^{-7}$ W/m$^{2}$ is reached at around 03:25 UT. The panels (from left to right) show the solar corona in the 211\AA\ EUV filter taken by the \textit{Atmospheric Imaging Assembly} (AIA; \citealt{2012lemen_AIA}) on--board of SDO, before, during and after the flare. Before the flare starts, the AR and the CH to the south can be well observed. As the flare progresses, the post--eruptive loop system and coronal dimming regions evolve related to the magnetic field restructuring and density depletion due to the erupting flux rope structure \citep[see \textit{e.g.},][]{2019dissauer}.

 \subsection{NLFFF Model}\label{subsec:nlfff}
 To qualitatively describe the configuration of the corona that leads to the non--radial evolution of the CME, we study the ambient magnetic field configuration of the Sun. We use synoptic vector magnetograms from the \textit{Heliospheric and Magnetic Imager} (HMI, \citealt{2012schou_HMI,2016couvidat_HMI}) on board SDO as input for a global nonlinear force--free field (NLFFF) model. The method was originally proposed in Cartesian geometry by \cite{2000wheatland}, but here we use the spherical optimization code developed by \cite{2007wiegelmann} and adjusted for the use of synoptic vector maps in \cite{2014tadesse}. As boundary condition we use a synoptic vector map for Carrington rotation 2111, which has been observed between June 05, 2011 and July 03, 2011.
 
\begin{figure} 
 \centerline{\includegraphics[width=1\textwidth,clip=]{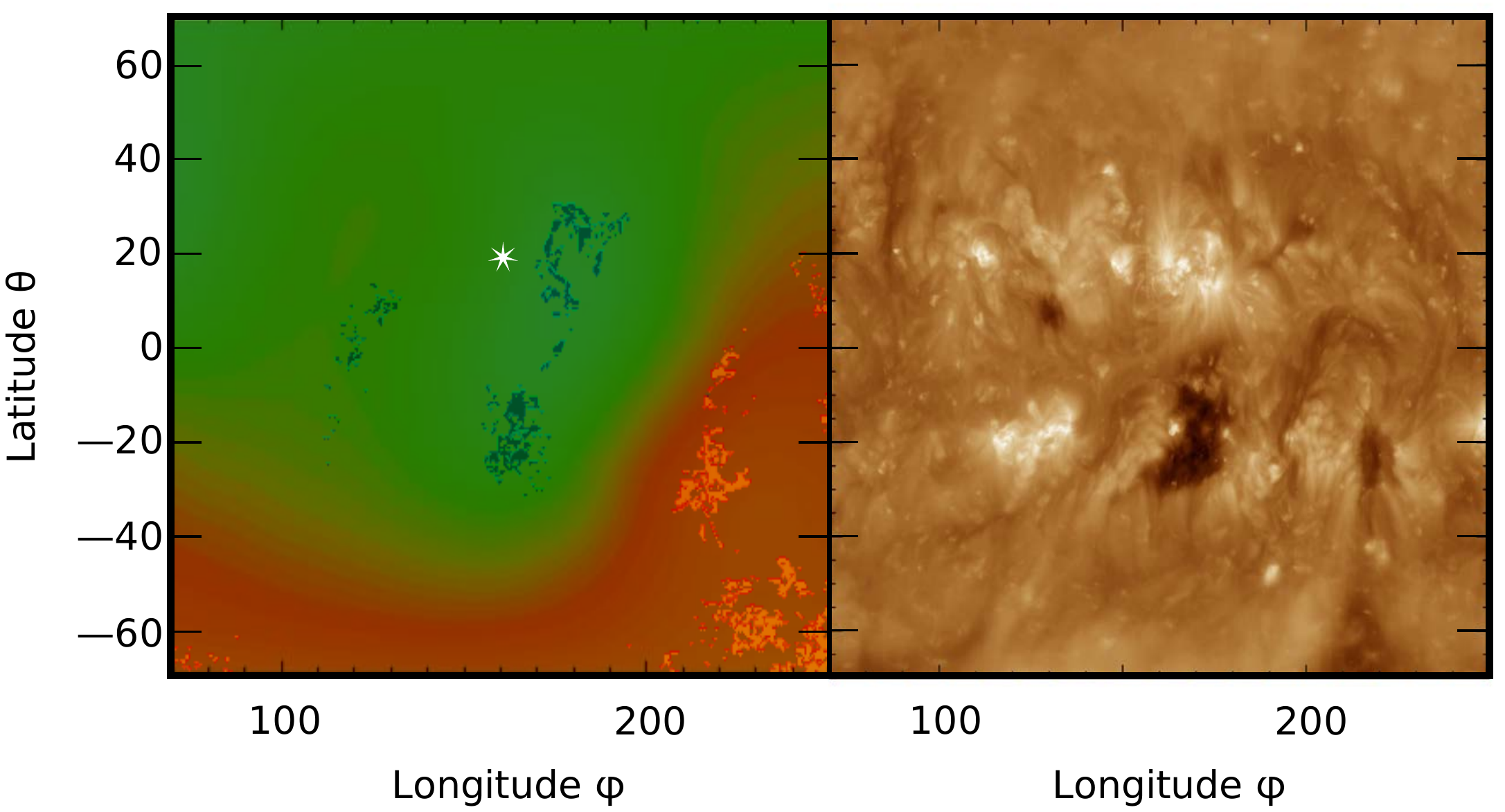}}
 \caption{\textit{Left}: CR2111, NLFFF--model at r=1Rs: Two--dimensional open field map $f(\theta,\phi)$ where $f = 1$ (\textit{yellow  underlying contours}) corresponds to positive and $f =-1$ (\textit{blue underlying contours}) to negative polarity footpoints of open field lines. $f = 0$ are areas hosting closed field lines. The NLFFF--model (Br) at $r=2.5$Rs is overlayed, with the \textit{red shaded area} representing positive polarity and \textit{green} representing negative polarity. The \textit{white star marks} the location of the AR. \textit{Right}: Synoptic image (CR2111) of the solar corona observed in the SDO/AIA $193$\AA\ filter for the same field of view.}\label{fig:NLFFF}
 \end{figure}

The left panel of Figure~\ref{fig:NLFFF} shows the resulting two--dimensional open field map at 1R$_{\odot}$ ($f(\theta,\phi)$) where yellow underlying contours corresponds to positive and blue underlaying contours to negative polarity footpoints of open field lines. Overlayed is the NLFFF--model (Br) at a surface height of $r=2.5~$R$_{\odot}$ with red representing positive polarity and green representing negative polarity. We see that the northern hemisphere has a negative polarity and vice versa the southern hemisphere has a positive polarity. Near the flux center of the flare we find open field footpoints to thesouth which are well observed in EUV as the CH (right panel of Figure~\ref{fig:NLFFF}). The open field footpoints to the west, indicative of the extension of the CH, might be outshined as a consequence of nearby bright loops, and are therefore not well visible as dark region in EUV. The identified open field footpoints most likely form a magnetic potential barrier towards south and west, that the CME cannot easily cross due to the \textit{frozen--in} condition in the coronal plasma. However, towards the north and east the CME can freely expand.

 \subsection{Coronal Dimmings and CME Launch}\label{subsec:dimming}
After establishing the conditions of the global solar magnetic configuration around the AR of interest, we analyze the evolution of coronal dimmings associated with the CME \textit{footpoints} anchored at the surface. Coronal dimmings are regions of strongly reduced emission observed in EUV and SXRs \citep{1996hudson,1997sterling,1998thompson,2000thompson}. They are a signature of the density depletion that is caused by the plasma exoansion and evacuation during the early CME eruption. Bipolar coronal dimmings are generally interpreted to represent the footprints of CMEs in the low corona (\textit{e.g.} \citealt{2000thompson,2019dissauer}) and can be used as proxies for studying the initial CME behavior. 

In order to properly track the dimming regions, we use a thresholding technique applied on logarithmic base--ratio EUV images. This also allows us to identify so--called secondary dimmings, which are mapping the overlying magnetic field that is expanding and erupting. They are of special interest, since they mostly reflect the propagation direction of the CME in its early evolution phase. To capture the full extent of coronal dimmings over time, we derive cumulative dimming masks. These masks contain all dimming pixels that are detected below a certain threshold and over a given time range. In this way different parts of the dimming that may grow and recover, \textit{e.g.} due to the associated flare, are also included in the detection. For further details on the method we refer to \cite{2018dissauer, 2019dissauer}.

\begin{figure} 
 \centerline{\includegraphics[width=1\textwidth,clip=]{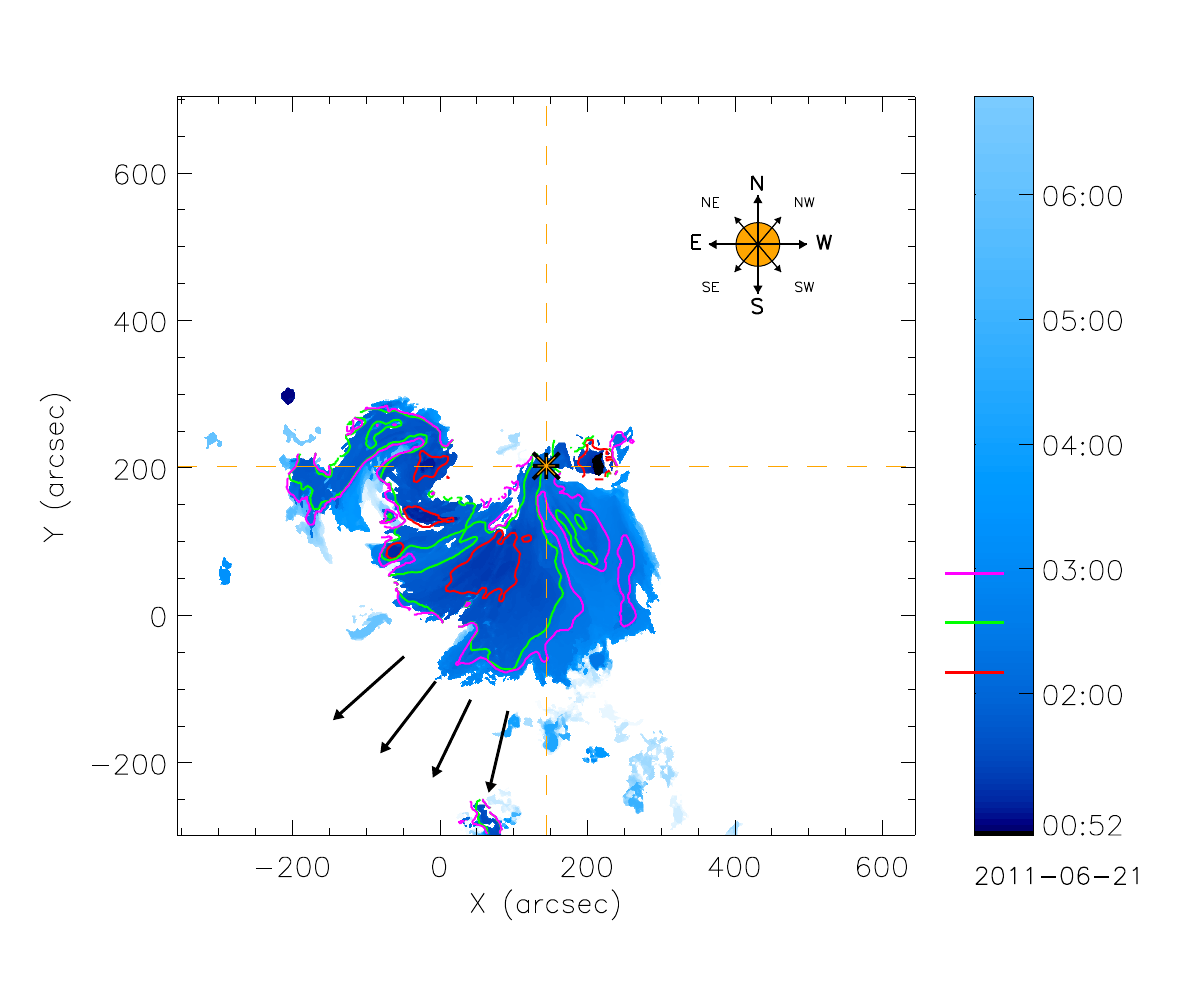}}
 \caption{Evolution of the coronal dimming caused by the CME. Each pixel is color--coded by the time of its first detection, where darker pixel represent an earlier detection time than lighter ones. The contours represent the size of the dimming region at three timesteps (as indicated by the \textit{colored lines in the colorbar}). The \textit{contours} are at 02:10 UT, 02:34 UT and 02:58 UT, which represent the impulsive evolution phase of the CME (up to 1 R$_{\odot}$ above the solar surface). The \textit{arrows} indicate the major evolution direction of the dimming. The image is centered at the flux center of the flare (\textit{yellow asterisks}).} \label{fig:dimming}
 \end{figure}

Figure~\ref{fig:dimming} shows the evolution of the CME associated dimming region, where each pixel is color--coded by the time of its first detection. Dark blue pixels are detected earlier than light blue pixels, respectively. The three contours represent the time evolution of the dimming region during its main impulsive phase, while the CME is propagating below $2~$R$_{\odot}$, at 02:10 UT(red), 02:34 UT (green), and 02:58 UT (magenta), respectively. The dimming mainly grows/spreads towards the south--east and south until 03:00 UT (marked as black arrows) in the reference frame of the flux center of the flare (yellow asterisk). This clearly shows that the CME was launched from the southern part of the AR, evolving into the direction of the CH. The eastward and south--westward spread may be caused by the CME lateral expansion. 

\begin{figure} 
 \centerline{\includegraphics[width=1\textwidth,clip=]{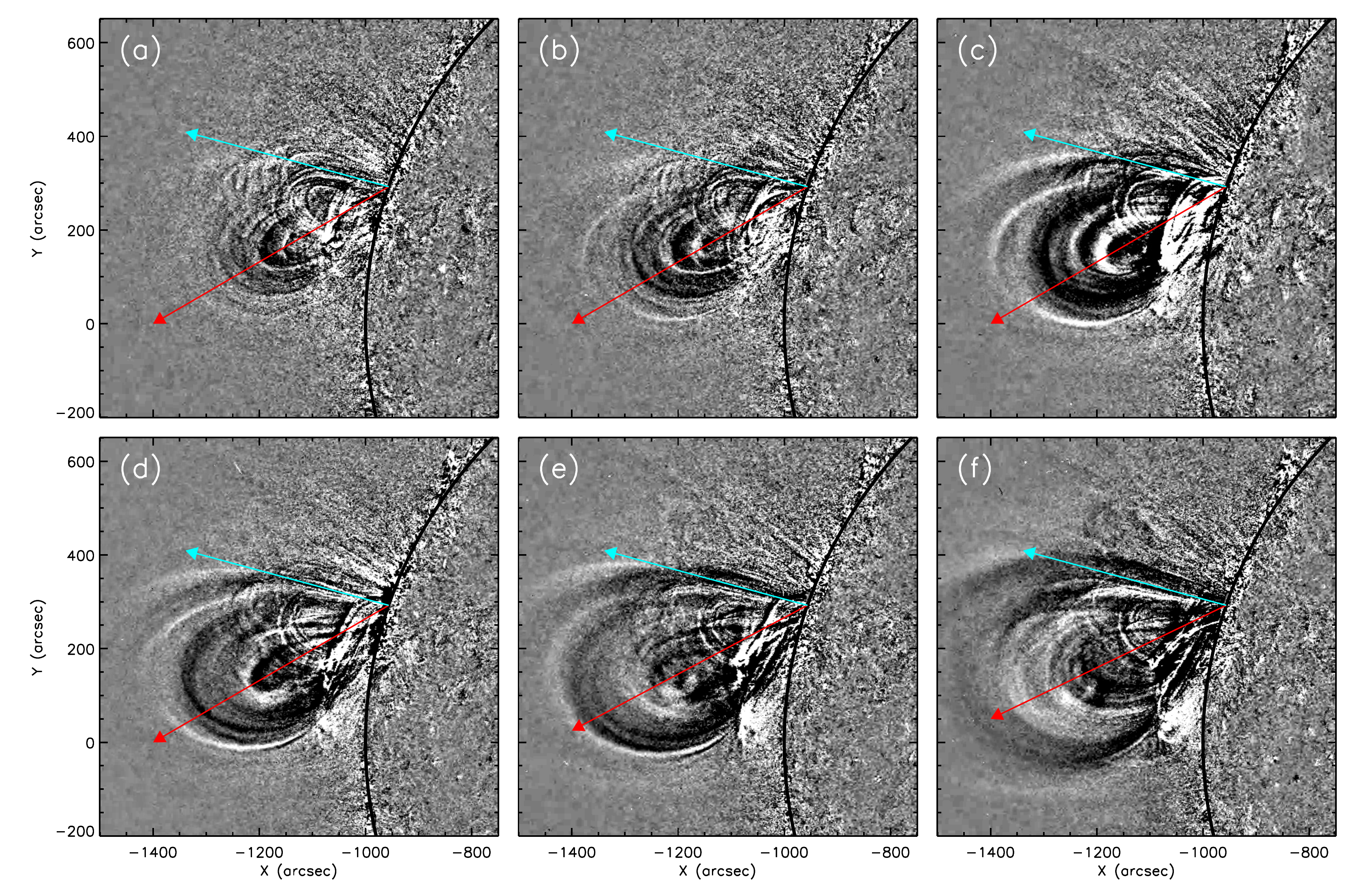}}
 \caption{Series of running difference images of the CME early evolution observed by EUVI/STEREO--A in 195\AA. The images are from 01:58 UT (a), 02:03 UT (b), 02:10 UT (c), 02:15 UT (d), 02:20 UT (e) and 02:25 UT (f) and show a clear southward propagation of the CME (\textit{red arrows}) in the first panels (a--d). In the last two panels a slowing down of the southward motion can be seen. The \textit{cyan arrows} indicate the radial direction.}\label{fig:evolution}
 \end{figure}

Figure~\ref{fig:evolution} gives the stereoscopic limb view of the CME using STEREO--A EUVI 195\AA\ running--difference images. We observe a clear southward propagation direction of the CME during its launch, which is in agreement with the early evolution of the associated coronal dimming. The red arrows indicate the observed propagation direction of the CME apex in the lower corona, which was estimated visually. It follows the trend shown in the latitudinal propagation profile of Figure~\ref{fig:forecat}. Following the CME evolution further on, one can clearly see that the CME gets deflected towards north over the distance range $1.3-3$ R$_{\odot}$, indicating an interaction with the open field from the CH.

 \subsection{ForeCAT and GCS}\label{subsec:forecat_gcs}
To further study the propagation of the CME we use two modeling approaches, the Forecasting a Coronal mass ejection's Altered Trajectory (ForeCAT; \citealt{2013kay,2015kay}) and the Graduated Cylindrical Shell (GCS; \citealt{2006thernisien,2009thernisien,2011thernisien}) model.
ForeCAT uses a Potential Field Source Surface model (PFSS; \textit{e.g.} see \citealt{1968schatten}) of the solar magnetic field to calculate the propagation, deflection and expansion of a CME on the basis of global magnetic pressure and tension. For further information on the model we refer to \cite{2015kay}.

Using the GCS model, we reconstruct the CME flux rope (FR) for 9 timesteps between 02:30 UT and 04:45 UT using COR1 and COR2 images from the SECCHI suite on both STEREO spacecraft in combination with LASCO C2 and C3 images \citep{1995brueckner}. For the GCS reconstruction, which was fitted to best represent the white--light features, of consecutive time steps, we derived large changes for latitude and longitude. This clearly shows the non--radial ejection and further deflection behavior of the CME due to its interaction with the open field of the HSS in the corona. Table~\ref{tab:gcs} shows the fitted GCS parameters.

\begin{sidewaystable}
\caption{GCS fitted parameters.}\label{tab:gcs}
\begin{tabular}{c|c|c|c|c|c|c|c}
\hline
Date &  Stonyhurst Longitude & Latitude & Tilt & Height & Aspect Ratio & Half Angle & Observational Data\\ 
 DD-MM-YYYY HH:MM:SS &  Degree & Degree & Degree  & $R_{\odot}$ &  &  Degree &    \\ \hline
21-06-2011 02:39:09	& $-20$ & $7$ & $-5$ & $2.3$ & $0.25$ & $20$ &COR1 (STA + STB)\\
21-06-2011 02:54:09 & $-20$ &$	8.5	$&$	-20	$&$ 2.95 $&$ 0.30 $&$ 15 $&	 COR1 (STA + STB)\\
21-06-2011 03:08:09 &$	-15 $&$	9 $&$ 	-30 $&$	3.7	$&$ 0.35 $&$	15	$&COR1 (STA + STB)\\
21-06-2011 03:23:09	& $ -15$	&$	9	$&$	-20	$&$ 5.1	$&$ 0.35 $&$	50 $&		C2, COR2 (STA + STB)\\
21-06-2011 03:39:09	&$ -15	$&$	10	$&$	-20	$&$ 6.5	$&$ 0.35 $&$	50 $&		C2, COR2 (STA + STB)\\
21-06-2011 03:54:09	&$ -15	$&$	12	$&$	-25	$&$ 8.0 $&$	0.40 $&$	50	$&	C2, COR2 (STA + STB) \\
21-06-2011 04:08:09	&$ -15	$&$	12	$&$	-20	$&$ 9.1 $&$	0.40 $&$	50	$&	C2, COR2 (STA + STB)\\
21-06-2011 04:23:09 &$	-15	$&$	15	$&$	-15	$&$ 10.5 $&$	0.40 $&$	40	$&	C3, COR2 (STA + STB)\\
21-06-2011 04:39:09	& $-15	$&$	15	$&$	-25	$&$ 12.0 $&$	0.40 $&	$40$ &		C3, COR2 (STA + STB)\\
\hline
\end{tabular}
\end{sidewaystable}

\subsubsection*{Initial CME propagation results from ForeCat and GCS}
\begin{figure} 
 \centerline{\includegraphics[width=1\textwidth,clip=]{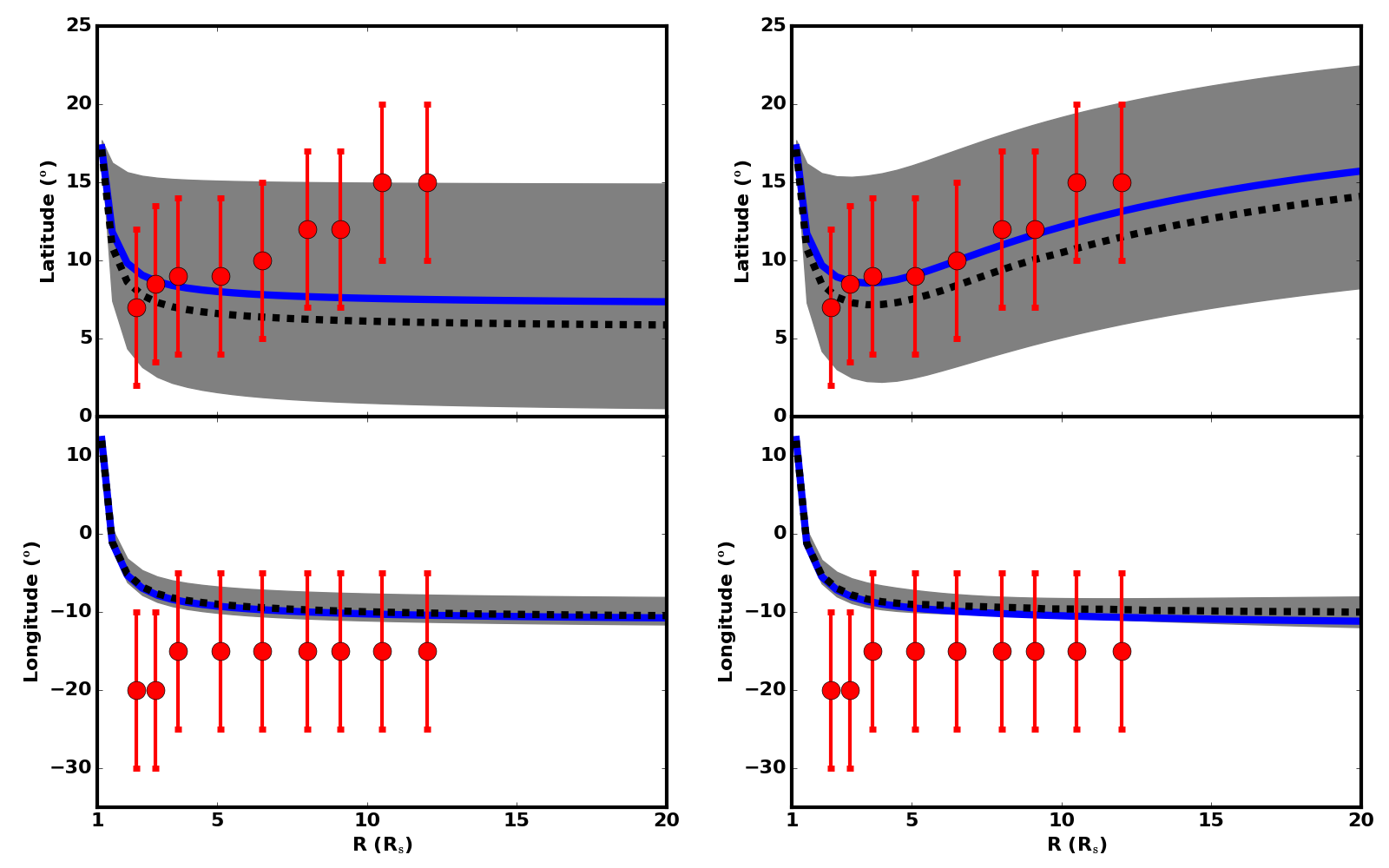}}
 \caption{Results of ForeCAT ensemble modeling in comparison to the GCS flux rope parameters. The top panels show the latitude and the bottom ones the longitude (Stonyhurst) of the CME flux rope. The \textit{blue line} shows the seed value of the ensemble. The \textit{dashed line} represents the median values, and the \textit{dark grey regions} are one standard deviation about that. The \textit{red dots} are the GCS values. The left panels show the results of the standard ForeCAT model, whereas in the right panels we included lateral overexpansion of the CME as well as artificial scaling of the magnetic field to simulate the compression of the HSS.}\label{fig:forecat}
 \end{figure}

Figure~\ref{fig:forecat} shows results of ForeCAT ensemble modeling \citep{2018kay} in comparison to the GCS FR parameters. The two top panels show the latitude of the CME axis and the two bottom panels show the longitude in Stonyhurst coordinates. We run an ensemble of 100 ForeCAT simulations with slight variations in the initial CME position and orientation. The blue line shows the \textit{seed} value of the ensemble -- the initial value that determines the center of the ensemble parameter range and our best guess at the true initial position and orientation. The dashed lines represent the median values, and the dark grey regions are one standard deviation about that. The red dots are the GCS values with error bars of $\pm5\arcdeg$ for latitude and $\pm10\arcdeg$ for longitude. The left panels show the results of the standard ForeCAT model, whereas in the right panels we included lateral overexpansion of the CME as well as artificial scaling of the magnetic field to simulate the compression of the field lines of the HSS. To simulate the compression we scaled the magnetic field strength in the direction of the CH by a scaling factor $S = 1 + (R - R_{0}) / 0.15 R_{\odot}$ which was estimated empirically, where R$_{0}$ is the initial height of the CME nose. 

We find the model CME to propagate from the northern hemisphere southwards and from west to east. After the initial phase the model suggests a near constant propagation direction at a latitude of $\sim7\arcdeg$ north and a longitude of $\sim10\arcdeg$ east. Starting at a height of about $2.3$ R$_{\odot}$ GCS reconstructions are available. The GCS model parameters are very similar to the ForeCAT results. There is a slight eastward offset in the longitude but within the error bars. Only in the latitude we do derive a significant difference between the two results: the GCS reconstruction shows the values continuously rising (northward motion of the CME due to interaction with the HSS) which we cannot derive from the results of the standard ForeCAT model (Figure~\ref{fig:forecat}, left panel), that uses a static background magnetic field to model the deflection of a CME. The interaction of a CME and a HSS however involves dynamic processes in which, among others, the open magnetic field structure of the HSS will be compressed and the magnetic pressure increases to a point where the CME can no longer expand or propagate in that direction, which could lead to a change in the CME's trajectory. We see this northward motion is reproduced when we include the overexpression and external compression in ForeCAT.

\subsubsection*{Flux Rope Evolution using GCS results}
 Using the last GCS reconstruction at 04:45 UT on June 21, and the near 1 au measured \textit{in situ} signatures of the magnetic ejecta (duration, average speed) we estimate the expansion factor of the FR in interplanetary space (IP). Assuming self--similar expansion at an arbitrary rate constrained by the initial (GCS) and final size (\textit{in situ}). We neglect the effects of the CME--HSS interaction in IP space on the CME trajectory and shape due to low plasma densities and magnetic fields.  Based on the GCS fitting results, we estimate that the initial FR radius at $13R_{\odot}$ is $3.7R_{\odot}$. From the \textit{in situ} measurement at 1 au, we estimate the FR radius to be $31R_{\odot}$.  The FR radius we estimate is the cross section along the Sun--Earth line to be able to investigate the geoeffective part of the magnetic structure. We note that this is the length along the observational path which means that it may not correctly represent the FR radius. The FR radius may be misrepresented in particular when the cross section is a skimming trajectory along the edge or a trajectory parallel to the FR axis. As we cannot clearly define the trajectory of the spacecraft through the magnetic structure (\textit{e.g.} force--free flux tube fitting results are inconclusive due to the low magnetic field strength in the magnetic ejecta), we assume that the observational path is a reasonable estimate of the FR radius but we also have to consider that the uncertainties are large. Using a power--law equation for the increase of FR size with heliospheric distance given by \cite{2018dumbovic} adapted from a more general expression by \cite{2008demoulin_solphys} and constrained by the derived GCS and \textit{in situ} results we estimate the power--law index $n_{A}$, \textit{i.e.}\, the expansion factor to $n_{A}=0.51_{+0.14}^{-0.13}$. We note that this resulting value lies on the lower end of the range obtained by statistical studies  \citep[\textit{e.g.}, see][]{2012gulisano,2015wangy} and indicates a relative slow increase in FR size or that the \textit{in situ} spacecraft passed through only a smaller part of the FR.
 
 Additionally, assuming a constant axial magnetic flux we can estimate the drop--rate of the central magnetic field strength, assuming that it follows a power--law behaviour with the power--law index $n_{B}$ \citep[see][]{2018dumbovic}. We expect $n_{B}$ to be $1.02_{+0.28}^{-0.26}$, which again is on the lower end of the range observed in statistical studies \citep{2012gulisano} and indicates a quite slow drop of the magnetic field strength within the FR. With a magnetic field strength of $10$ nT at 1 au we estimate the inital magnetic field strength at $13R_{\odot}$ to be $0.002$ G and the toroidal magnetic flux at 1 au to be $\phi=1.63_{+0.85}^{-0.63} \cdot 10^{20}$ Mx. This value is at the lower end of the expected range of typical ICME fluxes of $10^{20}$ to $10^{22}$ Mx for C to X--class events  \citep[\textit{e.g.},][]{2000devore,2007qiu,2015wangy,2017temmer-precond}. With such small expansion, we would expect the B field at 1 AU to be high (since the CME did not expand much) but in fact, B$_{1 \mathrm{au}} ~ 10$ nT, which is low. So either the initial field was very low to begin with and the \textit{in situ} spacecraft only encountered part of the FR or $n_{A}$ does not reflect $n_{B}$.

 \subsection{CME Kinematics in the Sun--Earth line}\label{subsec:kinematics}
After analyzing the initial solar configuration, the launch and the start of the CME--HSS interaction using observations and models, as next step we investigate the propagation of the CME in the direction of Earth. We derived the height--time profile of the CME from multiple sources. We manually tracked the CME front in the equatorial plane using EUVI, COR1, and COR2 image data separately from STEREO--A and --B spacecraft. In addition, we obtained the height--time profile from the intersection of the front of the GCS reconstructed shell with the Sun--Earth line.

\begin{figure} 
 \centerline{\includegraphics[width=0.9\textwidth,clip=]{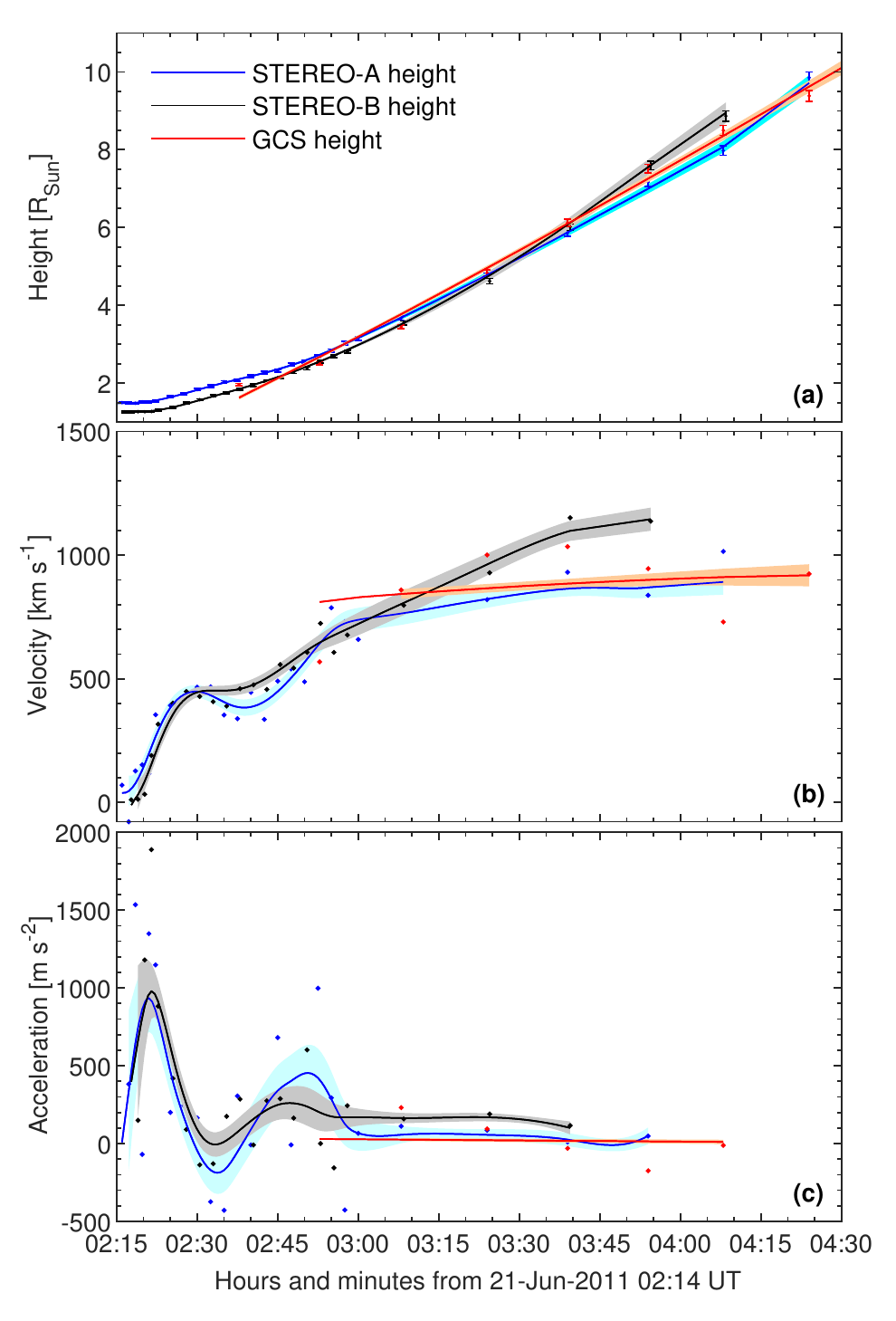}}
 \caption{CME kinematics up to a height of $10$ R$_{\odot}$. From \textit{top to bottom}: height, velocity, acceleration as function of time. The dots are the measured points and direct time derivatives, the solid lines are the fits to the measurements and the time derivatives of these fits. The different colors represent the different sources from which the kinematics were obtained: STEREO--A (\textit{blue}), STEREO--B (\textit{black}) and GCS (\textit{red}). The \textit{shadowed areas} represent the uncertainties.}\label{fig:kinematics}
 \end{figure}

Figure~\ref{fig:kinematics} shows the derived CME kinematics (top to bottom: height--time, velocity and acceleration profile). The dots are the measurements (height--time) and their direct numerical time derivatives (velocity, acceleration). To obtain robust estimates of the corresponding velocity and acceleration profiles, we first smooth the height--time curves and then derive the first and second time derivatives. The smoothing algorithm is based on the method presented in \cite{2017Podladchikova}, extended toward non--equidistant data.  The algorithm optimizes between two criteria in order to find a balance between data fidelity, \textit{i.e.}, the closeness of the approximating curve to the data, and smoothness of the approximating curve. From the acceleration profiles obtained in this way, we then interpolate to equidistant data points based on minimization of the second derivatives, and reconstruct the corresponding velocities and height profiles by integration (solid lines). The estimation errors of kinematic profiles are obtained by representing the  reconstructed CME height, velocity and acceleration as an explicit function of original CME height--time data in the assumption that STEREO--A and STEREO--B height errors are $1.5\%$ of height, and $3\%$ of height for GCS data. In the height--time diagram we derive that the CME is propagating from east (closer to STEREO--A) towards west (getting closer to STEREO--B) because of the initial offset between the black and blue line which cross later. The GCS model results are in good agreement with the direct measurements. The fit to the GCS height--time profile was done assuming that the acceleration has already subsided.

The velocity and acceleration profiles show the impulsive phase around 02:20 UT followed by a slower acceleration which subsides over time. Beyond a height of about $3~$R$_{\odot}$ the acceleration phase is finished and the CME propagates outwards at a speed of $800-1000$ km s$^{-1}$ which then seems to adjust to the speed of the surrounding HSS. Near Earth, the \textit{in situ} measured speed of the magnetic ejecta was $\sim600$ km s$^{-1}$ which was roughly equal to the speed of the HSS (see Figure~\ref{fig:in-situ}).

\section{Effects on the Magnetosphere} \label{sec:geomag}
\begin{figure} 
 \centerline{\includegraphics[width=1\textwidth,clip=]{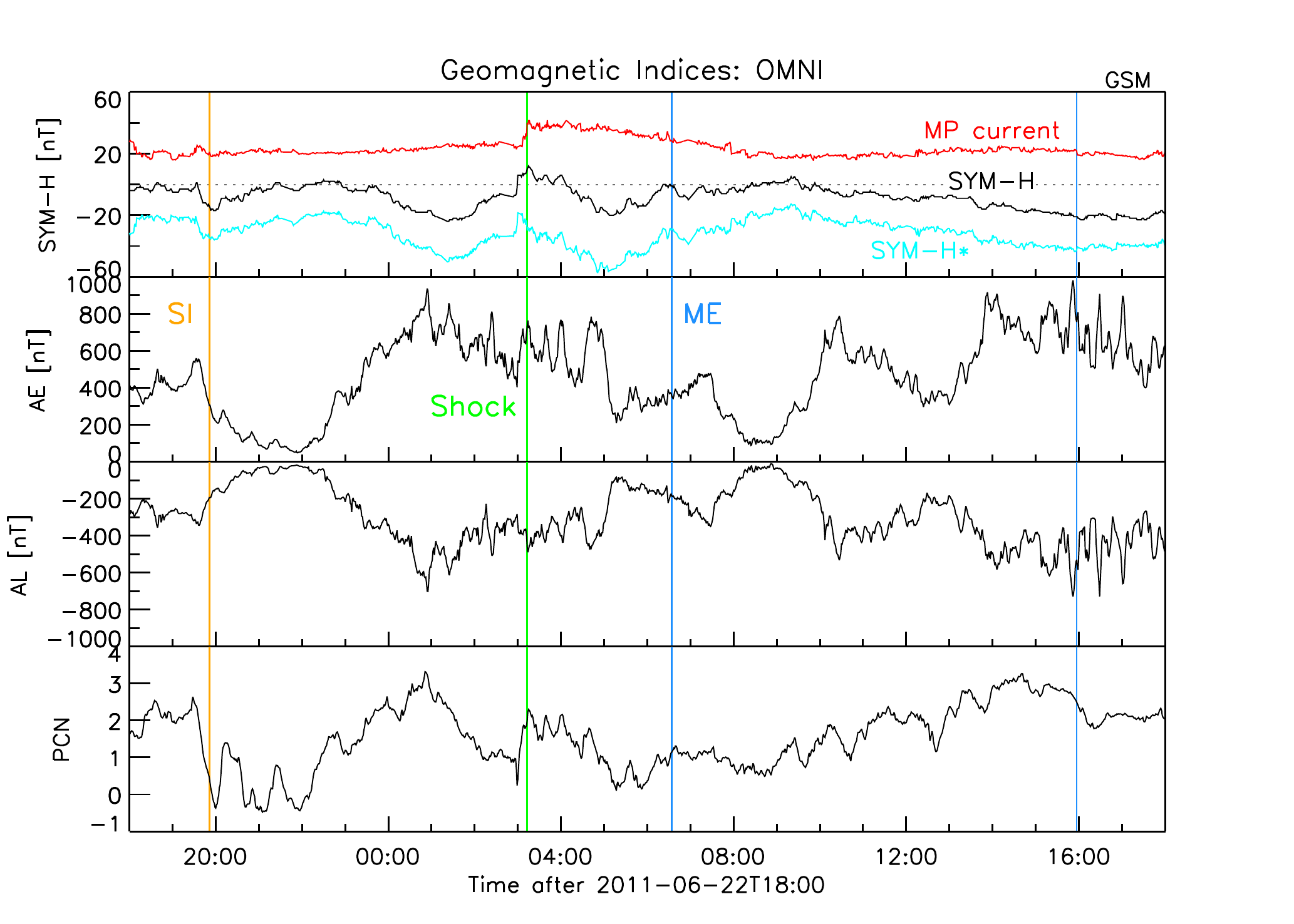}}
 \caption{Geomagnetic indices for the same time interval and including the same vertical guidelines as shown in Figure~\ref{fig:in-situ}. From \textit{top to bottom}: SYM--H (\textit{black}) including the magnetopause currents (\textit{red}) and the resulting corrected SYM--H* (\textit{cyan}), the Auroral Electrojet indices AE and AL and the Polar Cap Magnetic index (PCN). }\label{fig:geomag}
 \end{figure}

Finally we analyze the geomagnetic response. In Figure~\ref{fig:geomag}, we show the sym--H storm index, the auroral electrojet AE and AL indices and the PCN--north index. The sym--H panel also shows the correction due to the magnetopause currents (blue trace). The sym--H index shows fairly quiet conditions. Interestingly, correcting for the contribution of the magnetopause currents (sym--H*) almost doubles the effects, leading to a moderate storm at the high speed stream before the shock. Basically this is the effect of the relatively high dynamic pressure (see Figure~\ref{fig:in-situ}). Another moderate storm is caused by the CME sheath.

The AL and AE indices show signatures of a substorm at about 01 UT, June 23. At this time the ionospheric convection shows an increase (PCN index).  This convection enhancement is likely being contributed from the nightside source of substorm activity \citep{2012sandholt}. Some intermittent substorm activity is evident at the end of the interval.

\begin{figure} 
 \centerline{\includegraphics[width=1\textwidth,clip=]{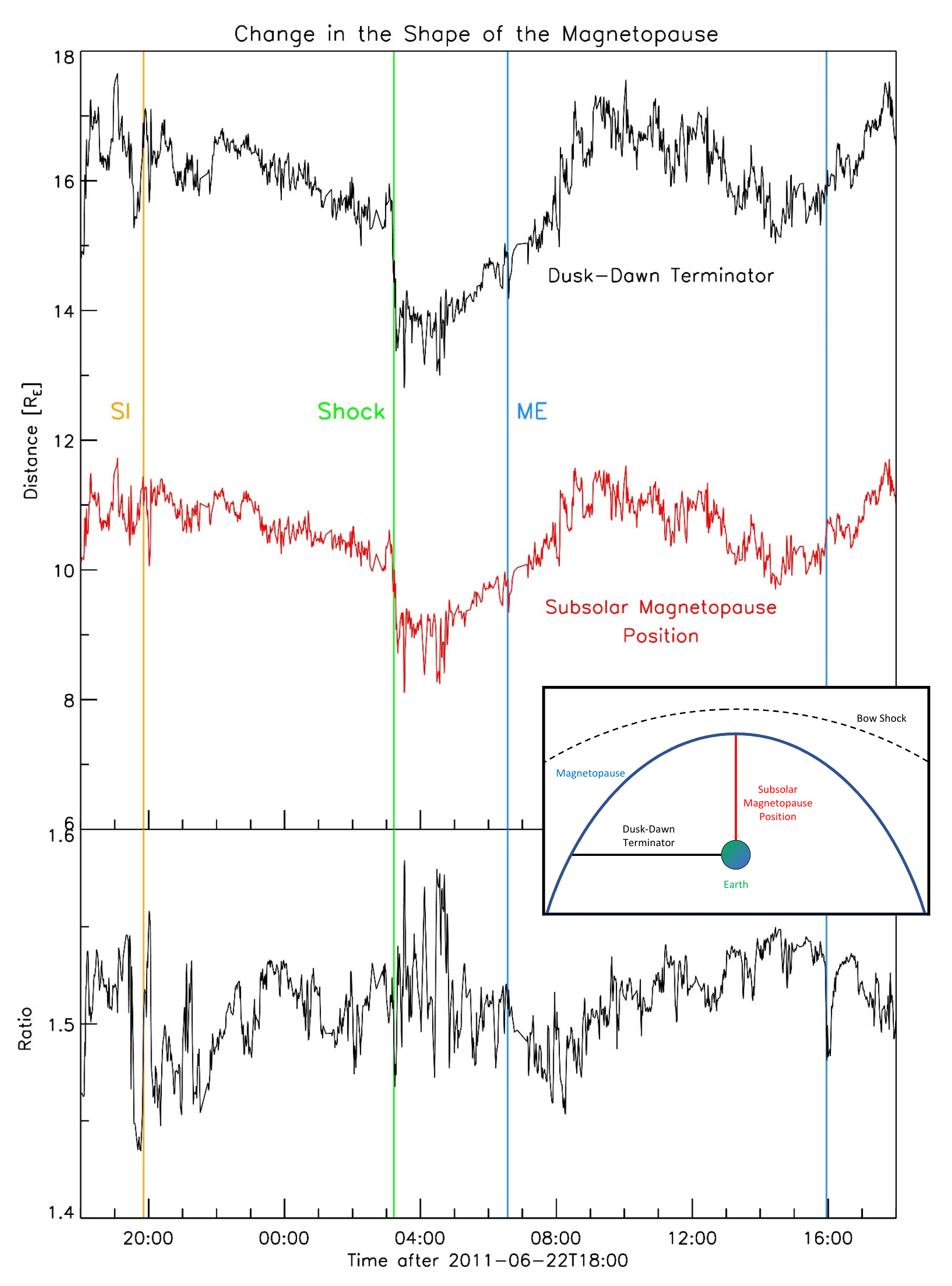}}
 \caption{Temporal profile of the magnetopause shape for the same time interval and including the same vertical guidelines as shown in Figure~\ref{fig:in-situ}. In the top panel the dusk--dawn terminator (\textit{black}) and the subsolar magnetopause position (\textit{red}) are shown. The bottom panel shows their ratio. The inset shows a cartoon demonstrating the shape of Earth's magnetosphere in regards to the dusk-dawn terminator and the subsolar magnetopause positon.}\label{fig:magshape}
 \end{figure}

As Figure~\ref{fig:in-situ} shows, the dynamic pressure is higher than typical (\textit{i.e.} 2 nPa). In addition, the B$_{\mathrm{z}}$ component oscillates with large amplitudes (about 5 nT) in the sheath region of the CME. These two factors are bound to have an effect on the magnetopause (MP) shape when we consider the \cite{1998shue} model which includes both the compression/rarefaction due to the dynamic pressure ($P_{\mathrm{dyn}}$) as well as the erosion due to the negative B$_{\mathrm{z}}$ component of the magnetic field. This result is confirmed Figure~\ref{fig:magshape}, which shows from top to bottom the temporal profile of subsolar magnetopause position, the dawn--dusk terminator and their ratio, \textit{i.e.} the aspect ratio of the magnetosphere.  
The subsolar distance (red) is closer to Earth than typical (\textit{i.e.} 12 R$_{\mathrm{E}}$). A large Earthward shift occurs at shock arrival. Thereafter, the magnetosphere generally expands slowly. The aspect ratio is typically 1.5. However, it changes significantly during the CME sheath passage visualizing the flaring of the magnetosphere. We conclude that the major effects were those produced by the sheath region.

 \section{Discussion}\label{sec:disc}
 
 In this paper, we present a detailed case study of a CME ejected on June 21, 2011 using multi--viewpoint remote sensing and \textit{in situ} observations supported by modeling to better understand and explain the intriguing \textit{in situ} solar wind signatures observed as a consequence of the CME interacting with a nearby CH. We find that the local and global magnetic field configuration, especially large--scale open magnetic structures such as CHs, have a major effect on the early propagation direction of the CME. The low--lying, local magnetic field configuration of the AR and its vicinity first leads to a non--radial ejection southwards. The global magnetic field structure, which in the close proximity is dominated by the open field lines of the CH, causes the subsequent deflection of the CME north--eastwards. 

From ForeCAT model results, we find that the open magnetic structures southwards and to the west of the AR seem to exert magnetic pressure on the CME, preventing it from propagating and expanding in these directions. The ForeCAT ensemble results show a larger spread in latitude than longitude, this is because the ensemble runs are based on how well the initial position within an active region can be identified. The spread in latitude and longitude in the corona is simply a result of how sensitive the model is to the precise initial position, and in this case the magnetic forces consistently give a strong eastward deflection but the latitudinal motion is slightly more varied. The resulting propagation direction is also supported by the NLFFF model results. \cite{2006cremades,2009gopalswamy,2013makela} found that there is a correlation between a CME's direction and the properties of nearby CHs (distance, area and mean magnetic field strength). For the CME under study, we find that the direction in which the CME is ejected is mainly affected by the configuration of the local magnetic field at low heights and the position of the flux rope (as estimated from EUV observations) rather than the nearby CH. The initial propagation of the CME to the south can be well observed off--limb in STB EUV images. The on--disk observations from the evolution of the coronal dimming associated with CME show the same behavior and can therefore be used as a proxy to derive the CME propagation direction \citep[see also][]{2019dissauer}. A similar conclusion is drawn by \cite{2007mandrini}, who analyzed the coronal dimming of the X17 flare event on October 28, 2003. 

Starting at a height of $\sim 1.3$ R$_{\odot}$ effects of the open field of the CH on the CME are clearly revealed. As the CME runs into the open field of the CH, the CME's further expansion and propagation in the direction of the open field is prohibited, resulting in a deflection towards north and east. This can be interpreted such that the southward motion is actually stopped and reversed by the compressed magnetic field of the HSS. While in the ForeCAT model this occurs as a net northward motion of the entire structure, in reality, where the CME is not forced to maintain a uniform shape, it could manifest as an asymmetric expansion in the northward direction. Separating these effects requires comparison of the external deflection forces and the internal forces that maintain a coherent CME structure, which is beyond the scope of this work. \cite{2011wang} defined three groups of deflected CMEs: asymmetrical expansion, non--radial ejection, and deflected propagation. Our event can  be classified as a combination of two different processes. Firstly, the CMEs initial direction deviates from the radial direction and can be classified as a non--radial eruption (second group). And secondly, due to the presence of an ambient magnetic structure the CME is deflected and as such belongs to the third group (deflected propagation).

The \textit{in situ} measured data near and at Earth show interesting signatures of the interaction between a SIR and a CME that occurred in the interplanetary space between the Sun and Earth. On June 22, 2011 at around 20:00 UT a clear stream interface signature can be identified, followed on June 23, 2011 at 03:15 UT by the HSS that also reveals a clear shock signature. The shock coincides with typical CME signatures starting at 06:30 UT, hence, can be attributed to be CME related. This however raises the question how that shock signature was produced. As shown in Figure~\ref{fig:in-situ}, the speed of the CME matches the bulk speed of the HSS before the shock arrival which suggests that the shock is not driven by the CME at this point. But the short duration of the shock--sheath region of about 3 hrs suggests otherwise. Due to the low compression of the magnetic field ($\sim1.5$) and of the density ($\sim2$) we can characterize the shock to have medium strength. We calculated the shock normal speed to be $\sim700$ km s$^{-1}$ and found that the shock can be considered quasi--parallel.

Due to the inconclusive results in deriving the flux rope geometry using the Lundquist force--free flux rope model were are not able to reliably determine the part of the structure which the spacecraft intersected. However we can estimate a trajectory from the modeled size and the propagation direction. As such, the most plausible interpretation is that the spacecraft only skims the outer edge of the CME and therefore the higher speed component, which drives the shock, is not observed \textit{in situ}. This interpretation is consistent with the steadily decreasing magnetic strength profile (as opposed to peaking later). As shown before, due to the magnetic configuration and the interaction with the HSS, the CME does not follow a radial direction but deviates at least $30\arcdeg$ from it. This could explain the relative short standoff distance (\textit{e.g.}, shock--sheath duration) as the shock could still be driven at the apex but the measured bulk speed at the intersection trajectory is equal to the speed of the HSS. In other words, the shock at the spacecraft crossing is being driven but not from the part of the CME at this trajectory rather from the plasma streaming from the apex. 
 
Based on these results, we suspect that the CME is engulfed by the HSS during most of its propagation to 1 au, which is consistent with the results from of the flux rope evolution. The flux tube expansion factor is with values of measured $n_{A}=0.51_{+0.14}^{-0.13}$ at the lower end of the range obtained by statistical studies \citep[see][]{2000devore,2012gulisano,2007qiu,2017temmer-precond}. Possible interpretations are that the FRs expansion was hindered and/or that only part of the magnetic structure is encountered by the spacecraft. We suspect both to have happened in this case.

The effects of this interaction on the Earth's magnetosphere are weak--to--moderate but show some surprising features such as a strong compression of the magnetorsphere and significant flaring during the CME sheath passage. We find some normal storm as well as substorm activity caused by the CIR. Due to the unusual high dynamic pressure of the shock, for a weak CME, the subsolar position of the magnetopause is decreased up to $27\%$ during the shock arrival, which causes significant magnetopause currents. This causes the storm activity to be considered moderate.

\section{Summary and Conclusion} \label{sec:summary}
We studied the origin of a peculiar \textit{in situ} signature measured at 1 au, caused by an Earth directed ICME and its interaction with a nearby HSS with the aim to adress two major questions:  How such a unique \textit{in situ} signature could be formed and whether the observed shock is still driven. Using a combination of observations and modeling efforts we were able to create a consistent interpretation of the event answering these questions:

\begin{itemize}
\item The CME is launched non--radially in south--eastern direction due to the configuration of the local magnetic field. This is reflected by the evolution of the dimming regions, the CME loops low in the corona (as seen in EUV observations) and the ForeCAT results.

\item At a height after $1.3$ R$_{\odot}$ the CME runs into the HSS which slows down the southward propagation until it is reversed and the CME is propagating northward. This is indicated by both the GCS measurements as well as the ForeCAT model. However, to be consistent with the GCS results, the ForeCAT model input requests an artificially scaled background field and a wider CME (overexpansion of CME cross section). With this the ForeCAT model mimics the effects of compression of the magnetic field due to the CH, that compresses the CME and deflects it away from it.

\item Due to deflection and the HSS ``wrapping'' around the CME, we measure a relatively small FR size and possible slow expansion (owing to the estimated low--value $n_{A}=0.51_{+0.14}^{-0.13}$).

\item The short standoff distance as well as the shock signature may be interpreted such as that the CME, which is engulfed by the HSS, only skims the spacecraft so its high speed part might have been missed. This interpretation is also consistent with the shape of the magnetic field profile  which is steadily decreasing and shows no clear peak as expected from a central flux rope crossing. Due to the magnetic configuration and the interaction with the HSS, the CME does not follow a radial direction but deviates at least $30\arcdeg$ from it causing this geometric effect in the measurements.

\end{itemize}

 65972 

%
 \begin{acks}
The SDO and STEREO image data and the WIND, ACE and STEREO \textit{in situ} data is available by courtesy of NASA and the respective science teams. We also thank the OMNI website at Goddard Space Flight Center, USA for a significant portion of the data used. S.G.H., M.T., K.D. and A.M.V. acknowledge funding by the Austrian Space Applications Programme of the Austrian Research Promotion Agency FFG (ASAP-13 859729 SWAMI, ASAP-11 4900217 CORDIM and ASAP-14 865972 SSCME). S.J.H.\,acknowledges support from the JungforscherInnenfonds der Steierm\"arkischen Sparkassen. C.J.F. acknowledges support from NASA Wind NNX16AO04G and STEREO Quadrature grant. N.L. and C.J.F. acknowledge support from NSF grant AGS-1435785. F.C. acknowledges the support by FEDER/Ministerio de Ciencia, Innovaci\'on y Universidades – Agencia Estatal de Investigaci\'on/Proyecto (ESP2017-88436-R) and Predoctoral Research Grants 2016-MCIU/FSE (BES-2016-077267). T.W. acknowledges DLR--grant 50 OC 1701 and DFG-grant WI3211/4-1. We thank the anonymous referee for constructive comments, which helped to improve the manuscript. Disclosure of Potential Conflicts of Interest: The authors declare that they have no conflicts of interest.
\end{acks}

%


%
%
%
%

\end{article} 
\end{document}